\documentclass[aps,prb,superscriptaddress,twocolumn,reprint,floatfix]{revtex4-2}
\usepackage{amsfonts}
\usepackage{amstext}
\usepackage{amssymb}
\usepackage{amsmath}
\usepackage{bm}
\usepackage[utf8]{inputenc}
\usepackage{graphicx}
\usepackage{graphics}
\usepackage{subfigure}
\usepackage{color,xcolor}

\usepackage{hyperref}
\hypersetup{colorlinks=true,citecolor=blue,linkcolor=blue,filecolor=blue,urlcolor=blue,pdfstartview=FitH,pdfpagemode=UseNone}
\usepackage{tikz}

\bibpunct{[}{]}{,}{n}{}{}

\begin{document}

\title{
Breakdown of topological protection due to non-magnetic edge disorder in \\ two-dimensional materials in the
Quantum Spin Hall phase}

\author{Leandro R. F. Lima}
\affiliation{Departamento de F\'{\i}sica, Instituto de Ci\^encias Exatas, Universidade Federal Rural do Rio de Janeiro, 
23897-000 Serop\'edica - RJ, Brazil}

\author{Caio Lewenkopf}
\affiliation{Instituto de F\'isica, Universidade Federal Fluminense, 24210-346 Niter\'oi, RJ, Brazil}
\affiliation{Max Planck Institute for the Physics of Complex Systems, 01187 Dresden, Germany}

\date{\today}

\begin{abstract}
We study the suppression of the conductance quantization in quantum spin Hall systems by a combined effect of 
electronic interactions and edge disorder,  that is ubiquitous in exfoliated and CVD grown 2D materials. 
We show that the interplay between the electronic localized states due to edge defects and electron-electron interactions gives rise to local magnetic moments, that break time-reversal symmetry and the topological protection of the edge states in 2D topological systems.
Our results suggest that edge disorder leads to small deviations of a perfect quantized conductance in short samples 
and to a strong conductance suppression in long ones.
Our analysis is based on on the Kane-Mele model, an unrestricted Hubbard mean field Hamiltonian and on a self-consistent recursive Green's 
functions technique to calculate the transport quantities.
\end{abstract}

\maketitle


{\it Introduction.--} 
The study of topological phenomena has grown enormously over the last years in condensed matter and material sciences, with significant impact in both fundamental and applied research \cite{Hasan2010,Qi2011,Ren2016,Culcer2020}. 
Of particular interest are two-dimensional (2D) topological insulators (TIs) characterized by robust edge states with a helical spin texture. These systems, also called quantum spin Hall (QSH) systems \cite{Hasan2010,Qi2011}, are promising platforms for transistor and spintronic applications \cite{Culcer2020}.
Theory predicts that QSH phases require, among other properties, a strong spin-orbit (SO) interaction.
The later can be intrinsic, like in inverted band semiconductor heterostructures \cite{Qi2011} and in a variety of 2D materials \cite{Marrazzo2019}, or extrinsic, generated by adatom doping \cite{Weeks2011} or proximity effects \cite{Avsar2020}. Accordingly, experiments reported QSH realizations in semiconductor quantum wells \cite{Konig2007,Roth2009,Gusev2011,Knez2014,Du2015}, 2D crystals \cite{Reis2017,Li2017,Fei2017,Wu2018}, and graphene with adsorbed clusters \cite{Hatsuda2018}.

Time-reversal symmetry and momentum-spin locking make the edge states robust against disorder, preventing backscattering and causing conductance quantization, ${\cal G}_0 = 2e^2/h$.
There are successful observations of localized edge states \cite{Roth2009,Nowack2013,Suzuki2013} and spin polarization \cite{Brune2012} in 2DTIs. 
However, the unexpected general experimentally determined
finite deviations from ${\cal G}_0$ in small systems and the conductance suppression in larger samples remain as a long standing and important puzzle 
\cite{Wu2018,Lunczer2019,Hsu2021}. 

The proposed backscattering mechanisms in 2DTIs can be divided into two main categories: interedge hybridization and intraedge spin-flip scattering processes.
Since the edge states typically have a penetration depth $\xi$ that is much smaller than the experimental sample widths ${\cal W}$, interedge hybridization 
is usually discarded. However, recent studies speculate that interface roughness in semiconductor heterostructures leads to chiral disorder that can create percolating paths enabling interedge scattering \cite{Lunczer2019}. 
In turn, since the magnetic impurities are rare in molecular beam epitaxy grown semiconductors as well as in exfoliated 2D materials, the simplest mechanism for spin-flip scattering to explain the lack of topological protection is also ruled out.  
This motivated several studies to explore a variety of ingenious mechanisms that effectively break time-reversal symmetry, namely,  noise \cite{Vayrynen2018}, edge reconstruction \cite{Wang2017}, Rashba SO interaction \cite{Strom2010}, phonons \cite{Budich2012}, nuclear spins \cite{Hsu2018,Bagrov2019}, charge puddles \cite{Vayrynen2013}, scattering processes due to adatoms \cite{Santos2018}, to name a few. 
Some of those give a temperature dependence at odds with the experimental findings \cite{Knez2014,Wu2018} and, more importantly, most are only suited for semiconductor heterostructures \cite{Konig2007,Roth2009,Gusev2011,Knez2014}. To the best of our knowledge, this is the first study to propose a breakdown of topological protection specific to 2D crystals. 

The combination of localization and electron-electron ($e$-$e$) interactions can also give rise to local magnetic moments. 
This feature is quite general and has been extensively studied in 2D materials, in particular the properties of vacancy induced localized states \cite{Dang2016, Zheng2018, Novelli2019} and of systems with zigzag terminated edges \cite{Wimmer2008,Pizzochero2021}.
Recently, Novelli and coauthors \cite{Novelli2019}  have shown that vacancies-induced magnetic moments destroy the topological protection. 
However, this effect occurs only within narrow energy resonances. Hence, despite being very insightful, this mechanism fails to explain the weak dependence of the conductance on gate potential observed in 2DTIs experiments \cite{Konig2007,Roth2009,Gusev2011,Knez2014,Du2015,Wu2018}.

In this Letter we put forward a new non-magnetic disorder mechanism to explain the breakdown of the topological protection in exfoliated and chemical vapor deposition (CVD) grown 2D materials. We show that edge disorder \cite{Mucciolo2009}, which is ubiquitous in exfoliated and CVD grown 2D materials, can lead to localization. We find that short sequences of zigzag edge terminations combined with $e$-$e$ interactions drive the formation of local magnetic moments that cause backscattering and destroy the conductance quantization in 2DTIs. We argue that the conductance suppression is small in short samples and can be large in longer ones, in line with experiments.


{\it Model --} We describe the system electronic properties within the topological gap using the Kane-Mele model Hamiltonian with a Hubbard term \cite{Kane2005,Novelli2019}
\begin{align}
\label{eq:modelH}
H = H_0 + H_{\rm SO} + H_U.
\end{align}
Here $H_0$ is the tight-binding Hamiltonian  
\begin{align}
\label{eq:modelH0}
H_0 =  -t \sum_{\langle i,j\rangle, \alpha} \left( c_{i\alpha}^\dagger c_{j\alpha}^{} + {\rm H.c.}\right)
\end{align}
where $c_{i\alpha}^\dagger (c^{}_{i\alpha})$ creates (annihilates) an electron of spin $\alpha$ at the honeycomb lattice site $i$ and $\langle i,j\rangle$
limits the hopping integrals to nearest neighbor sites. 

The second term 
describes the spin-orbit interaction due to adsorbed adatoms \cite{Weeks2011}
\begin{align}
\label{eq:modelHso}
H_{\rm SO} =  + i \lambda \!\sum_{\langle\!\langle i,j\rangle\!\rangle, \alpha \beta}  \!\nu^p_{ij} c_{i\alpha}^\dagger\sigma_{\alpha\beta}^z c^{}_{i\beta} 
\end{align}
where  ${\bm \sigma} = (\sigma^x, \sigma^y, \sigma^z) $ stand for $2 \times 2$ Pauli matrices in the spin space, $\langle\!\langle i,j\rangle\!\rangle$ 
restricts the sum to second neighbor sites, and $\lambda$ is the hopping integral energy. We assume that the adatoms are adsorbed at the 
so-called hollow positions  (centers of the hexagons) of the honeycomb lattice \cite{Weeks2011}, that we denote by $p$. 
Accordingly,  $\nu^p_{ij} = \pm 1$ distinguishes clockwise ($\nu^p_{ij} = 1$) and counterclockwise ($\nu^p_{ij} = -1$) hopping directions with respect to $p$ if the latter corresponds to an adsorbed adatom position, otherwise $\nu^p_{ij} = 0$.
The topological gap $\Delta_T$ is proportional to the adatom concentration, namely, $\Delta_T=6\sqrt{3}\lambda n_{\rm ad}$ \cite{Shevtsov2012}.
In the limit of $n_{\rm ad} = 1$ all $p$'s are filled and one recovers the original Kane-Mele model \cite{Kane2005}. 

Finally, we account for the $e$-$e$ interaction using an unrestricted Hartree-Fock approximation 
\footnote{
Some authors (see, for instance, \cite{Hsu2021} for a review) address the scenario where the $e$-$e$ interaction drives the helical edge modes into a Luttinger liquid. A consistent theory \cite{Hsu2021} requires a Luttinger liquid parameter  that corresponds to a strong $e$-$e$ interaction, which is not supported by experimental evidences, with the possible exception of InAs/GaSb quantum wells.
}
to the Hubbard Hamiltonian $H_U$, namely, 
\begin{align}
H_U^{\rm HF} \!=\! 
\frac{U}{2} \!\sum_{i,\alpha\beta} c_{i\alpha}^\dagger (n_i \mathbf{1}_{\alpha\beta}\! -\! {\bf m}_i \!\cdot\! {\bm \sigma}_{\alpha\beta} ) c_{i\beta} 
\!-\!\frac{U}{4} \! \sum_i (n_i^2 \!- \!|{\bf m}_i|^2),
\label{hamiltonianhf}
\end{align}
where $U$ represents the on-site (local) $e$-$e$ repulsive interaction, 
$\mathbf 1$ is the $2\times 2$ identity matrix, 
while
\begin{align}
n_i =  \sum_\alpha  \left\langle c_{i\alpha}^\dagger c^{}_{i\alpha} \right\rangle
\label{occupation}
\end{align}
is the mean electron occupation of the $i$-th site and 
\begin{equation}
{\bf m}_i =  \sum_{\alpha\beta} \left\langle c_{i\alpha}^\dagger  {\bm \sigma}_{\alpha\beta}^{} c^{}_{i\beta} \right\rangle
\label{magnetization}
\end{equation}
is related to the local electronic mean spin polarization, accordingly we refer to ${\bf m}_i$ as local magnetic moments. 


We consider a system of width $\mathcal W$ and length $\mathcal L$ with armchair edges along the transport direction, see Fig.~\ref{vacancy}(a).
Left ($L$) and right ($R$) contacts connect the system with source and drain reservoirs. 
For simplicity, we model the contacts by semi-infinite ribbons, with the same width as the central region and doped at $E_F\approx t$ to maximize the 
number of available propagating modes, mimicking metallic contacts.

{\it Methods --} 
We study the electronic transport 
using the nonequilibrium Green's-functions formalism (NEGF) \cite{Meir1992,Datta1995,Haug2008}. 
We use the spin-resolved linear conductance $\mathcal G_{\alpha\beta}$ as 
$\mathcal G_{\alpha\beta}(\mu) = (e^2/h)\int_{-\infty}^{\infty}\left(-\partial f_0/\partial E\right) \mathcal T_{\alpha\beta}(E)$ where $f_0(E)=[1+e^{(E-\mu)/k_BT}]^{-1}$ is the 
Fermi-Dirac distribution, $\mu$ is the equilibrium chemical potential and $T_{\alpha\beta}$ is the transmission coefficient  given by \cite{Meir1992} 
\begin{align}
	\mathcal T_{\alpha\beta}(E) = {\rm Tr} \left[\mathbf{\Gamma}_R(E) \mathbf{ G}_{\alpha\beta}^r(E)\mathbf{\Gamma}_L(E) \mathbf{ G}_{\beta\alpha}^a(E)\right].
\end{align}
Here $\mathbf G^{r}(E)=\left[E-\mathbf H-\mathbf\Sigma^{r}_R(E)-\mathbf\Sigma^{r}_L(E)\right]^{-1}$ and 
$\mathbf G^{a}(E)=\left[\mathbf G^{r}(E)\right]^\dagger$ are, respectively, the retarded and advanced Green's functions in the site representation and $\boldsymbol\Sigma^{r}_{R(L)}$ is the embedding self-energy  that depends on the retarded contact Green's functions and the coupling between the contacts $R(L)$ and the central region \cite{Datta1995,Haug2008}.
The $R$ and $L$-terminal linewidths are given by $\mathbf\Gamma_{R(L)}(E)=i[\mathbf\Sigma^r_{R(L)}-(\mathbf\Sigma^r_{R(L)})^\dagger]$. 
Since in our model  ${\mathcal T}_{\downarrow \uparrow} = {\mathcal T}_{\uparrow \downarrow} =0$, the total transmission is ${\mathcal T} = {\mathcal T}_{\uparrow \uparrow} + {\mathcal  T}_{\downarrow \downarrow}$.

We also analyze  the nonequilibrium local spin resolved conductance injected by the $R (L)$ terminal $\widetilde{\mathcal G}_{i\alpha,j\beta}^{R(L)}$, that is
given by $\widetilde{\mathcal G}_{i\alpha,j\beta}^{R(L)}(\mu) = (e^2/h)
\int_{-\infty}^{\infty}(-\partial f_0/\partial E) \widetilde{\mathcal T}_{i\alpha,j\beta}^{R(L)}(E)$  where  \cite{Lima2022}
\begin{align} 
\widetilde{\mathcal T}_{i\alpha,j\beta}^{R(L)}(E) = 2\ \text{Im}\left[\left(\mathbf G^r \mathbf\Gamma_{R(L)}\mathbf G^a\right)_{j\beta,i\alpha}H^{}_{i\alpha,j\beta}\right]
\end{align} 
is the local transmission of electrons flowing from site $j$ with spin $\beta$ to the site $i$ with spin $\alpha$. 
The spin resolved local current is obtained {using the local version of the Landauer-B\"utiker equation $\widetilde{\mathcal G}_{i\alpha,j\beta}^{R} V_R+\widetilde{\mathcal G}_{i\alpha,j\beta}^{L} V_L$, where $V_R$ and $V_L$ are the terminal voltages \cite{Lima2022}.

We calculate $\mathbf G^r$ using 
the recursive Green's-function (RGF) technique \cite{Mackinnon1985,Lewenkopf2013,Lima2018}, 
and $\mathbf \Gamma_{R,L}(E)$ by decimation \cite{LopezSancho1985,Lewenkopf2013}.
The system Green's function depends self-consistently on $n_i$ and ${\bf m}_i$ that we obtain using optimized methods (see Supplemental Materials \cite{supmat} for details).
For technical reasons \cite{supmat, Lima2016} we perform our calculations at finite temperature keeping $k_BT \ll \Delta_T$, in line with all situations of interest. 
Hence, in what follows we neglect thermal smearing effects and take $\mu=E_F$.

\begin{figure*}[tb]
	\includegraphics[width=2\columnwidth]{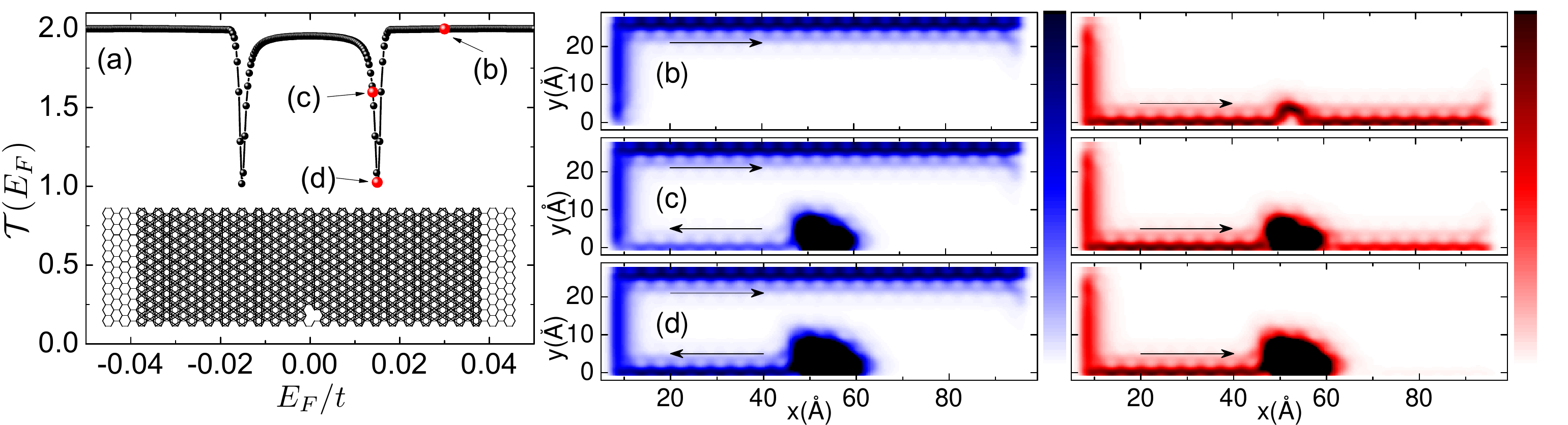}
	\caption{(a) Total transmission $\mathcal T$ as a function of $E_F/t$. Inset: Sketch of the system geometry with a vacancy close 
to the bottom edge.  Hexagons with lines connecting second neighbors bonds indicate the positions of the adatoms, 
here $n_{\rm ad} =1$. 
Local transmissions injected by the left lead ${\mathcal T}_{i\uparrow,j\uparrow}^L$ (blue) and ${\mathcal T}_{i\downarrow,j\downarrow}^L$  
(red) for (b) $E_F = 0.030t$, (c) $E_F =0.014t$, and (d) $E_F =0.015t$. The color intensity stands for the magnitude of ${\mathcal T}_{i\alpha,j\alpha}^L$
and the arrows indicate the electron flow direction. }
	\label{vacancy}
\end{figure*}

{\it Vacancy-induced magnetic moments.$-$}
%
The occurrence of vacancies in honeycomb lattices gives rise to quasi-localized states \cite{Lieb1989,Pereira2008,Nanda2012},
For a sufficiently strong $e$-$e$ interaction,  due to the Stoner instability, these localized states lead to the formation of 
local magnetic moments  \cite{Yazyev2010,Ugeda2010,Zhang2016}. 
For the Hubbard mean field approximation, it has been shown that any finite $U$ causes magnetization \cite{Kumazaki2007}.
Vacancy-induced magnetism has been recently proposed as a mechanism to explain the breakdown of the conductance quantization in TIs \cite{Novelli2019}. 
Here, we briefly review this setting and argue that this is hardly a suitable mechanism to explain the lack of conductance quantization 
observed in experiments. 

We consider a system with a vacancy near the bottom edge, see Fig.~\ref{vacancy}(a). 
We take $\mathcal W=27$\AA, that is sufficiently wide to show no numerical evidence of backscattering due to interedge hybridization.
We set $n_{\rm ad}= 1$ and $\lambda= 0.1t$, which leads to $\Delta_T=1.04t$, and analyze the electronic transport in a chemical potential window within the QSH phase. 
For these parameters we find enhanced backscattering for $U\gtrsim U^* \approx 0.34t$ \cite{supmat}. 
Without loss of generality, we consider $U=0.4t$.
Other values of $U \ge U^*$ do not change qualitatively our findings \cite{supmat}.

Figure~\ref{vacancy}(a) shows the computed total transmission ${\mathcal T}(E_F)$ within the QSH regime.
The $\mathcal T(E_F)$ is perfectly quantized outside the energy window comprised by two narrow antiresonance centered at $\pm E_r$.
At the antiresonance there is total backscattering, whereas for $E_F$ between antiresonances  ${\mathcal T}(E_F)$  is slightly suppressed with respect to ${\cal T} = 2$.

The involved scattering processes can be qualitatively understood by analyzing the local transmissions shown in Figs.~\ref{vacancy}(b) to (d).
For $E_F = 0.030 t$ the total transmission is quantized, the electrons injected by the $L$-terminal are transmitted by the helical edges states without backscattering,  see Fig.~\ref{vacancy}(b). For $E_F = 0.014t$ the transmission is no longer perfect. 
The local magnetic moment at the sites around the vacancy cause a spin-flip process enabling partial backscattering that reduces the $L\rightarrow R$ transmission of spin down electrons, see Fig.~\ref{vacancy}(c). This situation becomes more dramatic at the antiresonance peak $E_F = E_r = 0.015t$, for which the $L\rightarrow R$ transmission of spin down electrons is fully blocked, see Fig.~\ref{vacancy}(d).
 
The magnetization around the vacancy decays as a power-law \cite{Nanda2012,Miranda2016}, that explains the relative large regions 
over which the edge currents invert their direction of propagation, see Figs.~\ref{vacancy}(c) and (d).
Note that the edge states decay exponentially towards the bulk with a penetration depth $\xi = \hbar v_F/\Delta_T$, suppressing the spin-flip coupling as the vacancy is moved away form the edge. 

The observed features can be quantitatively described by an effective single-edge low energy model with a Hilbert space that consists of the vacancy-induced quasi-localized state $|0\rangle$ at $E=0$ \cite{Pereira2008} and helical edge modes ($|k,\uparrow\rangle$ and $|-k,\downarrow\rangle$). 
This simplification allows one to map the microscopic problem into the model of an edge state scattered by a short-range magnetic impurity proposed in Ref.~\cite{Dang2016}. 
The analytical solution \cite{Dang2016,Zheng2018} of the corresponding Lippmann-Schwinger equation gives a single-edge transmission that is very similar to that of Fig.~\ref{vacancy}(a). A detailed comparison between the microscopic and the effective model shows that a strong conductance suppression is always associated with resonance processes with narrow energy decay widths $\Gamma$, namely, $\Gamma/\Delta_T \ll 1$ \cite{supmat}.

We conclude that vacancies are unlikely to explain the breakdown of conductance quantization observed in experiments, since:
({\it i}) Vacancies have low concentrations in good quality samples.
({\it ii}) The model requires that the vacancies occur very close to the system edges (within the edge state penetration depth). 
({\it iii})  The conductance suppression occurs only over a very narrow chemical potential doping window, a feature that 
has not been observed in experiments.  Diluted distributed 
vacancies are hardly expected to modify this scenario.


{\it Wedge defect.$-$}
%
Local magnetic moments can also be originated by $e$-$e$ interactions in localized states 
due to the system edge terminations. This subject has been extensively investigated in graphene zigzag 
\cite{Son2006,Yazyev2010} and chiral \cite{Tao2011} nanoribbons. 
Interestingly, it has been theoretically shown  that a small sequence of zigzag links is sufficient to spin polarize the 
system for any finite $U$ \cite{Wimmer2008,Kumazaki2008}.
Let us study the simplest lattice edge geometry that leads to the formation of a local magnetic moment for our model Hamiltonian,
Eq.~\eqref{eq:modelH}: a wedge or V-shaped defect. 
We note that a similar kind of edge defect has been recently observed experimentally in high precision bottom up 
nanoribbon graphene synthesis \cite{Pizzochero2021}.

Now we take $ n_{\rm ad}= 0.5, \lambda= 0.1t$, and keep $\mathcal W = 27 $\AA, that is sufficiently 
wide to prevent interedge scattering and exhibit perfect conductance quantization for $U=0$. 
Figure~\ref{fig:vshapetransmission} shows the total transmission $\mathcal T$ as a function of $E_F/t$ 
for $U =t$. 
For small zigzag wedges, $\ell = 3$ and $\ell = 5$ links long, the transmission is already no longer quantized and the deviation from perfect 
transmission shows a tendency to increase with $E_F$. 
Remarkably, there is a sharp transition for zigzag wedges with $\ell = 6$ links and longer:
The transmission along the edge with the V-shape defect becomes zero over a large fraction of the topological gap energy window.

Here, the localized edge states are very different from the vacancy-induced ones and so is the local magnetization.
We find that the in-plane magnetic moments are strongly peaked and almost constant along the $\ell$ zigzag chain at the system edge and show a fast decay for sites with increasing distance to the edge \cite{supmat}. 
The local transmissions show that the spin-flip processes occur in the vicinity of the wedge defect, suggesting that $\delta w/\xi \agt 1$, were $\delta w$ is the depth of the wedge, is required to maximized the effect. For the parameters we consider, the latter inequality implies $\ell \agt 5$, consistent with Fig.~\ref{fig:vshapetransmission}.

\begin{figure}[tbp]
	\centering
	\includegraphics[width=0.9\columnwidth]{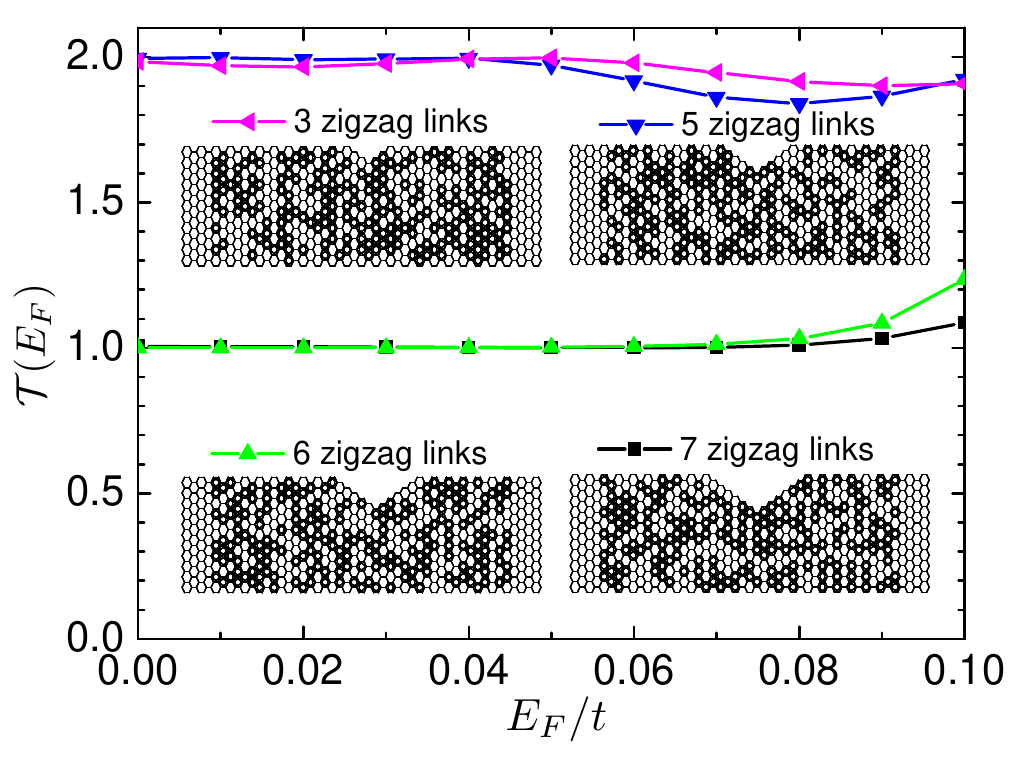} 
	\caption{Total transmission $\mathcal T(E_F)$ for wedge-shaped edge defects of different lengths. The inserts show wedge defects with $\ell = 3, 5, 6$, and 7 zigzag links. The darker sites correspond to the hexagons with adsorbed adatoms, in all cases $n_{\rm ad}=0.5$. }
	\label{fig:vshapetransmission}
\end{figure}

 {\it Rough edge disorder.--}
%
Let us now study a more realistic edge disorder model.
Both for lithographic and CVD synthesized samples, the system edges can be assumed as rough, with no clear crystallographic orientation. 
Nonetheless, experiments show \cite{Zhang2013,Leicht2014} that experimental samples do display sequences of zigzag and armchair links.
Below, we investigate the impact of edge roughness on the wedge-defect induced magnetic moments.  

The inset at the bottom of Fig.~\ref{fig:rough_edges}(a) shows a realization of rough edge disorder.
We consider a system with a single wedge defect and generate the edge disorder by randomly removing atoms from 
the two hexagonal rows at the system top edge. Next, we remove any atom forming dangling bonds.
We find that the edge roughness increases the edge states penetration depth $\xi$.
We increase the system width to $\mathcal W \approx 57$\AA ~to prevent inter-edge scattering.
The inset at the top of Fig.~\ref{fig:rough_edges}(a) shows the corresponding  $m_{ix}$ at $E_F=0$ and
indicates that the presence of edge roughness does not break the wedge defect induced magnetization.
As expected, we find that the magnetization of longer zigzag chains $\ell$ is stronger than that of sequences with a smaller $\ell$.

Figure~\ref{fig:rough_edges}(a) shows $\mathcal T (E_F)$ for the disorder realization presented in the inset. 
Due to edge roughness, backscattering is slightly suppressed as compared to Fig.~\ref{fig:vshapetransmission}, but the qualitative behavior is similar: 
The transmission through the top edge goes to zero over a wide Fermi energy window inside the topological gap, remaining finite for other valeus of $E_F$.
Figures~\ref{fig:rough_edges}(b) shows $\widetilde{\mathcal T}_{i\alpha,j\alpha}^{L}(E_F=0)$. 
{Spin-up electrons} flowing from $L$ to $R$ are scattered by the local magnetic moment [see upper inset of Fig.~\ref{fig:rough_edges}(a)] that flips their spins and force them to propagate back to the $L$ terminal, as displayed by {Figs.~\ref{fig:rough_edges}(b) and \ref{fig:rough_edges}(c) at $E_F=0$}. 
The propagation from $L$ to $R$ at the bottom edge has perfect transmission.   

\begin{figure}[th]
	\centering
	\includegraphics[width=0.95\columnwidth]{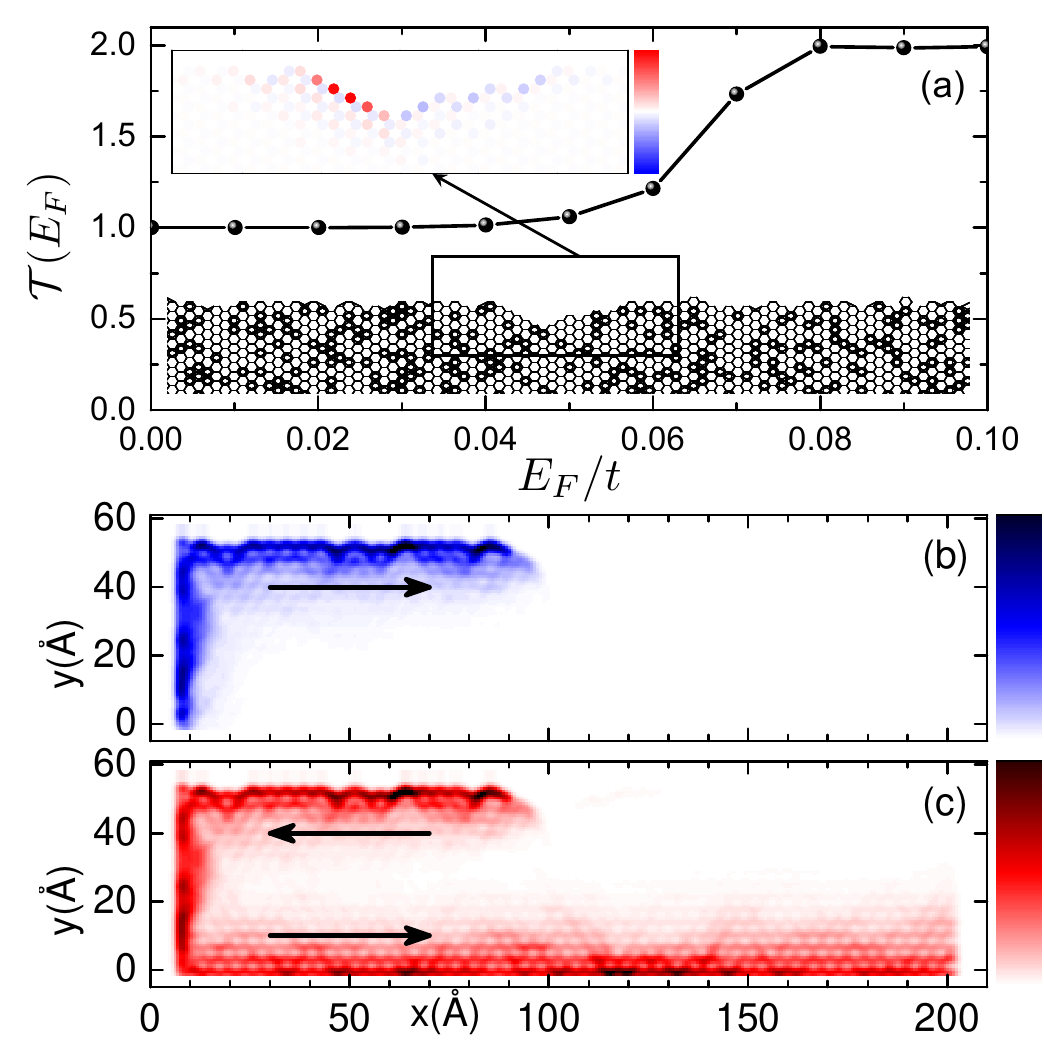}
	\caption{(a) Total transmission $\mathcal T(E_F)$ for the disorder realization and local moment shown the lower and upper insets. 
		Local transmission   $\widetilde{\mathcal T}_{i\alpha,j\alpha}^{L}(E_F=0)$ in arbitrary units for (b) for spin up ($\alpha = \uparrow$) and (c) spin down ($\alpha = \downarrow$)  electrons. }
	\label{fig:rough_edges}
\end{figure}

Clearly, the probability for the occurrence of long zigzag chains is smaller than that of short ones. 
The picture that emerges is: 
Systems with small ${\cal L}$ are dominated by short $\ell$ wedge defects that lead to small deviations from conductance quantization, $2 - {\cal T} \ll 1$. 
With increasing  ${\cal L}$,  wedge-defects with $\ell > 5$ are more likely to occur resulting in a strong suppression of ${\cal T}$.

{\it Summary.--} 
%
We have studied the suppression of the conductance quantization in exfoliated and CVD grown 2D materials in the QSH regime due to local magnetic moments caused by edge disorder, that is ubiquitous in such systems.
We find that the interplay between a single wedge defect and $e$-$e$ interactions can destroy topological protection 
over a large extension of the topological gap, either causing a small modification in the conductance quantization or 
strongly suppressing the conductance. We conjecture that the latter in more likely to occur in large samples.
We believe that this mechanism is quite general and applicable to a large variety of intrinsic 2D topological insulators as well as 
for extrinsic adatom doped and proximity 2DTIs.


The authors acknowledge the financial support of the Brazilian funding agencies CNPq, FAPERJ and the 
Brazilian Institute of Science and Technology (INCT) in Carbon Nanomaterials. 
The simulations were partially performed at 
NACAD-COPPE/Federal 
University of Rio de Janeiro, Brazil.

\bibliography{spin_hall_adatoms_disorder}

\end{document}